\documentstyle[prl,aps]{revtex}
\begin{document}
\title{Exponentially rapid decoherence of quantum chaotic systems}
\author{Arjendu K. Pattanayak and Paul Brumer}

\address{Chemical Physics Theory Group, \\ University of Toronto, Toronto,
Ontario, Canada M5S 3H6} 
\date{16th Sept 1997}
%%%%%%%%%%%%%%%%%%%%%%%%%%%%%%%%%%%%%%%%%%%%%%%%%%%%%%%%%%%%%%%%
\maketitle
\draft
\begin{abstract}
We use a recent result to show that the rate of loss of coherence of a 
quantum system increases with increasing system phase space structure and 
that a chaotic quantal system in the semiclassical limit decoheres 
exponentially with rate $2 \lambda_2$, where $\lambda_2$ is a generalized 
Lyapunov exponent. As a result, for example, the dephasing time for 
classically chaotic systems goes to infinity logarithmically with the 
temperature, in accord with recent experimental results. 
\end{abstract}
\pacs{PACS numbers: 05.45.+b,03.65.Sq,03.65.Bz,05.40+j}

Decoherence\cite{decoh,joos} which is, in general, the loss of interference 
effects in a quantal system due to its interaction with the 
environment\cite{leggett}, is important in fields that rely on the 
maintenance of delicate quantal coherence,
including the rapidly-growing areas of coherent control\cite{control},
quantum computation\cite{computation} and charge transport in mesoscopic
devices\cite{mohanty}. It also plays a central role in treatments of 
classical-quantum correspondence\cite{decoh,joos}.
Preliminary studies have suggested that systems which
are chaotic in the classical limit decohere rapidly\cite{zp,hu,tamesht}.
However, the rate of this decoherence and its dependence on system 
properties is unclear.
Here we derive the functional form for the rate of decoherence, provide an
explicit link between the rate and the extent of the system's phase space
structure and demonstrate that decoherence occurs exponentially, with a 
rate determined by a generalized Lyapunov exponent, for quantum systems 
with a chaotic classical analogue. In addition, we note that the decoherence 
rate is reduced for $\hbar$ sufficiently large, although in general the 
decoherence of a quantal system may be maximally rapid for a finite value of 
$\hbar$. Finally, we show that our results
explain the recent experimental result\cite{mohanty} that the decoherence
rate in mesoscopic devices saturates as the temperature $T\to 0$. All of 
these results follow from an analysis of decoherence in light of a recent 
approach\cite{ap_pb_2} which connects the exponentially fast growth of 
structure in the distribution dynamics of chaotic systems to a generalized 
Lyapunov exponent $\lambda_2$ of the classical dynamics in phase space. 

Consider the Wigner representation ($\equiv\rho^W$)\cite{groot} of a
quantal density matrix; it is suitably normalized (${\rm Tr}[\rho^W] =1$),
but is in general not a `pure state' ($ {\rm Tr}[(\rho^W)^2]\neq 1$). The
relevant decoherence equation for $\rho^W$, derived from the quantum system
evolving under the potential $V(q)$ and coupled to an external environment
through the position operator, is \cite{decoh,joos,zp,leggett}
\begin{equation}
\label{wigner}
{\partial \rho^W\over\partial t}=\{H,\rho^W\}+\ \sum_{n \geq
1}\frac{\hbar^{2n}(-1)^n}{2^{2n} (2n +1)!} \frac{\partial^{2n+1}}{\partial
q^{2n+1}}V(q) \; \frac{\partial^{2n+1}}{\partial p^{2n+1}}\rho^W 
+ D \frac{\partial ^2 }{\partial p^2}\rho^W. 
\end{equation}
The first term on the right of this equation is the Poisson bracket or
classical Liouville operator, which generates the classical evolution for
$\rho^W$; the terms in $\hbar$ add the quantal evolution, with
the expansion being formally valid for a potential $V(q)$ analytic in $q$.
The decoherence is contained in the term dependent on $D$; its precise form 
and parameter-dependence rests upon assumptions about the form of
the system-environment coupling and the spectrum of the environment. The
standard form\cite{decoh}, for example, has $D = 2\gamma M k_B T$ where
$M$ is the system mass, $T$ the temperature of the environment, $k_B$ is
the Boltzmann constant and $\gamma$ an unknown measure of the
system-environment coupling. Different forms for $D$ may also
obtain\cite{joos,hu_paz}, although $D$ always depends monotonically on the
coupling and the temperature of the environment. The particular form of the
coupling, $D\frac{\partial ^2 }{\partial p^2}\rho^W$ 
comes from the choice of coupling through position; coupling through
momentum would give $D \frac{\partial ^2 }{\partial q^2}\rho^W$.
For full generality it is appropriate to use 
$D\nabla^2\rho^W$ where $\nabla^2$ is the Laplacian operator in phase
space. By using the relationship $\frac{d \hat A}{dt} = \frac{\partial \hat
A}{\partial t} + \frac{1}{i\hbar}[\hat A, \hat H]$ for the time-evolution 
of an arbitrary operator $\hat A$, Eq.~(\ref{wigner}) can also be written as
\begin{equation}
{d \rho^W\over d t}= D\nabla^2\rho^W.
\label{total}
\end{equation}
We use the $\nabla^2$ form hereafter; the arguments are entirely unchanged
if either the momentum or the position coupling is used.

Our criterion for monitoring the degree of system decoherence is the
Renyi entropy\cite{decoh,schlogl} $S = \ln({\rm Tr}[(\rho^W)^2])$. This 
function maximizes at zero for a pure state and is a direct measure
of the degree of mixing of the quantum state\cite{decoh,xupei}. The 
time-dependence of $S$ thus provides information about the rate at which 
a ``coherent" quantum pure state is transformed into an ``incoherent" or 
statistical mixture of states. 
It is straightforward to show, using Eq.~(\ref{total}) that
\begin{equation}
{dS\over dt} = 2D\frac{{\rm Tr}[\rho^W \nabla^2\rho^W]} { {\rm Tr}
[(\rho^W)^2]} \equiv - 2D\chi^2\label{rate}.
\end{equation}
The quantity $\chi$ \cite{ap_pb_2} (which is in general time-dependent)
affords insight insofar as it is the root-mean-square Fourier radius of the 
distribution, and a measure of phase space structure. That is, if we Fourier 
expand any distribution $\rho$ as
$\rho(p,q) =\int\int d\mu\;d\nu\;e^{2\pi i(\mu p + \nu q)}
\tilde{\rho}(\mu,\nu)$ where the tilde now represents the
Fourier-transformed function, we get that 
\begin{equation}
\chi^2 =- \frac{{\rm Tr}[\rho \nabla^2\rho]}{{\rm Tr}[\rho^2]} ={\frac{{\rm
Tr}[\;|\nabla\rho|^{2}]}{{\rm Tr}[\rho^{2}]}} = \frac{4 \pi^2\int dk\;
k^2|\tilde{\rho}(k)|^2}{\int dk\;|\tilde{\rho}(k)|^2}
\end{equation}
where the second equality follows from integration by parts and where 
$k \equiv (\mu,\nu)$. Since higher Fourier modes are related to increased 
structure, $\chi$ measures the structure in the distribution, and 
Eq.~(\ref{rate}) shows that {\em the decoherence rate of a quantal system is 
directly proportional to the extent of system structure in phase 
space}\cite{footnote1}.
Note that since ${dS\over dt}$ is negative-definite, the system can not
``recohere"\cite{footnote4} and hence much of 
the decoherence is determined by the short-time behavior of $\chi(t)$.
Therefore, as long as the term in $D$ is not initially dominant, we can
make quantitative predictions of $dS/dt$ based upon the time-dependence
of the phase space structure from the Hamiltonian part of the evolution for
$\rho^W$\cite{footnote2}.

Consider first time-independent states (e.g., a mixture of Hamiltonian
eigenstates). To first order, i.e. before decoherence becomes significant, 
we can estimate $\chi^2$ from the initial $\rho^W$, and find that lower
energy states, with less structure, decohere more slowly. Thus, for example, 
for the $n$th eigenstate of the harmonic oscillator $\chi^2 \propto
n$ so that decoherence is proportional to $n$, explaining the numerical
results of Knight and Garraway\cite{knight}.
Hence we deduce that in general stationary distributions with lower
energy are more stable with respect to environmental perturbation.

For time-evolving distributions we need to consider Eq. (\ref{rate}).
For nonchaotic systems some insight results from a simple analysis since in 
this instance an adequate measure of the time-dependent $\chi^2$ is the 
time-averaged $\bar{\chi^2}$. For pure states of the harmonic oscillator for
example, working within the approximation that an initially pure state
remains so, this criterion yields that minimum uncertainty coherent state
are the most stable\cite{decoh}.

For chaotic systems insight obtains from examining the behavior of $\chi^2$ 
in the semiclassical limit. In this limit $\rho^W$ is known\cite{ap_pb_2} to 
behave like the classical result, at least for early times. Consider then the 
behavior of $\chi^2$ for a classically chaotic system. 
Let the equations of motion of a point in
phase space be $\dot {x} = f(x)$. The equations of motion for the
vectors in the tangent space are obtained by linearizing the dynamics $x
\to x + \varsigma$ around a fiducial trajectory to yield
\begin{equation}
\label{eq:Mij}
\frac{d {\varsigma}}{dt}= {\cal M}{\varsigma} 
\end{equation}
where the Jacobian matrix ${\cal M}$ has the elements $M_{ij}=
\frac{\partial f_j}{\partial x_i}$ at the point $x(t)$. It is
straightforward to show\cite{ap_pb_2} that 
\begin{equation}
\frac{d}{dt} {\nabla}\rho = - {\cal M}{\nabla}\rho, \label{nabla}
\end{equation}
that is, the equations for evolution of ${\varsigma}$ and ${\nabla}\rho$
are {\em identical} except for a minus sign, where $\nabla\rho$ is the 
phase-space gradient of the classical probability density $\rho$. This implies 
that the usual maximal Lyapunov exponent may be computed from the distribution 
gradient evaluated along a trajectory as 
\begin{equation}
\label{eq:lambda}
\lambda (x) = \lim_{t\to\infty}\frac{1}{t} \ln(|\nabla\rho(x(t))|)
\end{equation}
We now note that $\chi^2$ is the distribution average of this gradient. Thus,
adaptating the results of the standard thermodynamic approach for similar
averages which use $\varsigma$ instead of $\nabla\rho$\cite{schlogl}, we
have \cite{ap_pb_2} that 
\begin{equation}
\lim_{t \to \infty} \frac{1}{t}\ln(\chi) = \lambda_{2}. \label{eq:lambda_2}
\end{equation}
Here $\lambda_{2}$, which is independent of trajectories, is a generalized
Lyapunov exponent related to the ordinary Lyapunov exponent by $\lambda_2 =
\lambda + \zeta$, where $\zeta \geq 0$ measures the local phase-space
fluctuations in $\lambda$ (equivalent to a measure of the degree of
intermittency in chaos\cite{schlogl}). The quantity $\lambda_2$ has the 
particularly simple interpretation\cite{ap_pb_2} in that it measures the 
exponentially rapid increase of structure for a chaotically evolving 
distribution in a Hamiltonian system and is in this sense a fundamental 
exponent for chaos in distribution dynamics. From this treatment it is clear 
that $\chi(t)$ grows exponentially rapidly in a chaotic system as
\begin{equation}
\chi^2(t)= \chi^2(0)[\exp(2 \lambda_2 t) + C] \label{eq:exp_chi}.
\end{equation}
where $C$ in general fluctuates or grows slower than exponentially with time.

Thus, substituting Eq.~(\ref{eq:exp_chi}) in Eq.~(\ref{rate}) we get that 
the initial decoherence rate of a quantal system in the semiclassical limit, 
to first order in $D$ is
\begin{equation}
{dS\over dt} = - 2D\chi^2(0)[\exp(2\lambda_2t) + C],
\label{o_d2}
\end{equation}
an equation which displays a direct correlation between decoherence rates
and the generalized Lyapunov exponent of the underlying classical system.

Note that Eq.~(\ref{o_d2}) implies that in the limit of chaotic semiclassical 
behavior, all time-dependent quantum states decohere exponentially
rapidly and hence that there is only a marginal dependence on the initial
distribution. This is in contrast to non-chaotic systems where, as noted
above, the rate is strongly affected by the initial state.

We have numerically solved the discrete map version of Eq.~(\ref{wigner}), 
with the full $\nabla^2$ coupling to the environment, for a quantum chaotic 
Cat Map\cite{josh,ap_pb}, to explore the nature of $S(t)$ as a function of
$\alpha$ (a scaled $\hbar$\cite{josh,ap_pb}). Results for $\alpha = 10^{-1}$ 
to $10^{-5}$ are shown in Fig.~(1), with $D$ chosen at a typical value of 
$10^{-6}$. Several features are evident: (a) in the limit of
$\alpha\to 0$, $S$ decreases exponentially rapidly initially with a rate 
determined by $\lambda_2=\lambda=0.9624$\cite{footnote3}, consistent
with Eq.~(\ref{o_d2}), (b) the initial decoherence rate increases with $\alpha$ 
for small $\alpha$, with maximal falloff for an $\alpha \approx 0.001$, and 
(c) for $\alpha$ sufficiently large, the entropy decreases quite slowly.

Examination of the first quantum correction to Eq.~(\ref{o_d2}) provides
some insight into the observed dependence of $S(t)$ on $\alpha$. Note first 
that as $\hbar$ is increased away from $0$ and for small $t$, there is an 
{\em increase} in the amount of structure in the distribution. That is, one 
expects $\rho^W$ to adiabatically follow the classical distribution, but with 
added fringes\cite{berry} (interference structures). The added structure can 
be seen, for example, in comparing the contour maps (Figs.~(1-3), 
Ref.~\cite{ap_pb}) of the evolving distributions. From the first quantal term 
in Eq.~(\ref{wigner}), we can estimate that the quantum correction to an 
individual Fourier component $\tilde{\rho}(k)$  is $\tilde{\rho}(k) \to
\tilde\rho^W(k) \approx \tilde{\rho}(k)(1 + c'\hbar^2k^3)$ where $c'$
depends on the potential and includes all other constants. Thus, for an initial
distribution that is not very structured (vanishing support at large $k$),
$\rho^W$ is very close to $\rho$. However, even when the quantum terms are 
small, $\chi^2$ sums them over all the Fourier components, weighted by the 
factor $k^2$. A crude estimate\cite{ap_pb_u} then yields that the quantum 
corrected $\chi^2_q$ relates to the classical estimate $\chi^2_c$ as 
$\chi^2_q \approx \chi^2_c( 1+ c\hbar^2\chi_c^6)$. Hence the quantum
correction to $\chi^2$ can be quite large. 

This analysis of the contribution of an individual Fourier component to the 
enhancement of structure breaks down for $\hbar^2k^3 = {\cal O}(1)$; this 
happens rapidly, since for classically chaotic dynamics the support for the
distribution at larger $k$ values increases exponentially fast. Hence, 
higher-order contributions also have to be estimated, which is in general 
extremely difficult. By this time, however, the second quantal effect enters: 
Quantum distributions resist the growth of structure at scales smaller than 
$\hbar$\cite{berry} and the support of the distribution travels slowly 
across the $k\approx\frac{1}{\alpha}$ boundary in Fourier space. 
This is marked by\cite{ap_pb_2} a slow-down of the growth of $\chi$ at 
${\cal O}(\frac{1}{\alpha})$. The interplay between these two effects leads 
to a value of $\hbar$ where the quantum system decoheres maximally rapidly 
for early times. Quantum systems for $\hbar$ larger than this critical value 
are, however, stable against decoherence, as seen in Fig.~1.

A second interesting effect emerging from this treatment is the dependence of 
the decoherence rate on $D$. It has been recently observed\cite{mohanty} that 
the decoherence rate for quantal mesoscopic devices goes to zero slower than 
any power of the temperature $T$ as the systems are cooled to near-zero $T$. 
The measure used in these experiments is the dephasing time, which is 
monitored through the change in coherent quantal back-scattering effects in 
the device (weak localization)\cite{mohanty}. In our theoretical analysis, 
arguing that an increase in the degree of ``incoherent" mixing reduces this 
coherent back-scattering effect, a reasonable analogue of the dephasing time 
may be taken to be the time $t_R$ at which the system has ``sufficiently 
decohered", i.e when the entropy has fallen to some fixed fraction of its 
initial value $R = S(0) - S(t_R)=\ln(\frac{{\rm Tr}[(\rho^W(0))^2]}{{\rm Tr}
[(\rho^W(t_R))^2]}) = 1$, for example. 
Since these mesoscopic devices are disordered systems, with the 
electrons experiencing substantial chaotic scattering from impurities, 
Eq.~(\ref{o_d2}) is expected to apply in the semiclassical limit. If we now 
solve for the effect on $S(t)$ of the dominant exponential term from this 
equation, we get the expression
\begin{equation}
t_R =\frac{1}{2\lambda_2}\ln\bigg [\frac{R\lambda_2}{D\chi^2(0)} + 1\bigg].
\label{Dbehav}
\end{equation}
The $1/D$ term dominates as $T$, and hence $D, \to 0$ and indicates that, 
{\em independent} of the particular power-law dependence of $D$ on the 
temperature, the dephasing time goes to infinity {\em logarithmically} with 
vanishing temperature, i.e, slower than any power-law. Equation~(\ref{Dbehav}) 
is, in fact, able to reproduce the general shape of Figs.~(1) and (2) of 
Ref.~\cite{mohanty}, albeit with free parameters.  In addition to this 
logarithmic dependence on $D$ in the semiclassical limit, a different effect
shows up as the system becomes more quantal: Since (1) the decoherence rate 
grows as $D\chi^2$ and (2) the growth of structure slows down at 
$\chi = {\cal O}(\frac{1}{\hbar})$, the decoherence term has a distinctly 
greater impact if $\sqrt{D} \geq {\cal O}(\hbar)$. This behavior can be seen 
in the slow decrease in $S$ for $\alpha =10^{-2},10^{-1}$ in Fig.~1.

In summary, we have demonstrated that the decoherence rate of a quantal 
system coupled to the environment is governed by the degree of structure 
in the system as measured by the quantity $\chi$. This has enabled us to 
make various deductions about the stability of distributions under the action 
of the environment which generalize several recent\cite{decoh,zp,hu} 
discussions. We have shown that since the structure of a distribution 
increases exponentially fast for a classically chaotic system, distributions 
in such systems decohere exponentially fast in the semiclassical limit, with 
the exponent given by the quantity $\lambda_2$. This behavior is manifest, for
example, in the observation\cite{mohanty} that the dephasing time for 
semiclassical systems which are classical chaotic goes to infinity
logarithmically with the temperature.

We thank the Natural Sciences and Engineering Research Council and the 
U.S. Office of Naval Research for support of this research.
A.K.P. thanks Harold Baranger for useful discussions on Ref.~\cite{mohanty}
and Salman Habib for stimulating correspondence on an earlier version of 
this paper.

\begin{figure}[htbp]
\caption{Time-dependence of the degree of incoherent mixing of quantal 
distributions as measured by the change in entropy ($S(t)-S(0)$) for the 
chaotic quantal Cat Map coupled to the environment. The effect of approaching 
the classical limit is shown by varying $\alpha$ from $10^{-1}$ to $10^{-5}$, 
where $\alpha$ is a scaled $\hbar$. $D$, which is a measure of the coupling 
to the environment, is $10^{-6}$. See text for details.}
\end{figure}


\begin{thebibliography}{99}
\bibitem{footnote1} A different choice of coupling variable would lead to
the decoherence rate being dependent on some combination of
$Tr[\rho^W\frac{\partial ^2 }{\partial p^2}\rho^W]$ and
$Tr[\rho^W\frac{\partial ^2 }{\partial q^2}\rho^W]$ in general.
\bibitem{footnote4} This is a direct manifestation of the irreversibility of
Eq.~(1), which is an approximation to the full dynamics (See, e.g.
\cite{leggett,decoh}).
\bibitem{footnote2} This is equivalently a first-order approximation in $D$ 
or a short-time result.
\bibitem{footnote3} In systems where the local stretching rate is constant,
such as the Cat Map, $\lambda_2=\lambda$.
\end{thebibliography}
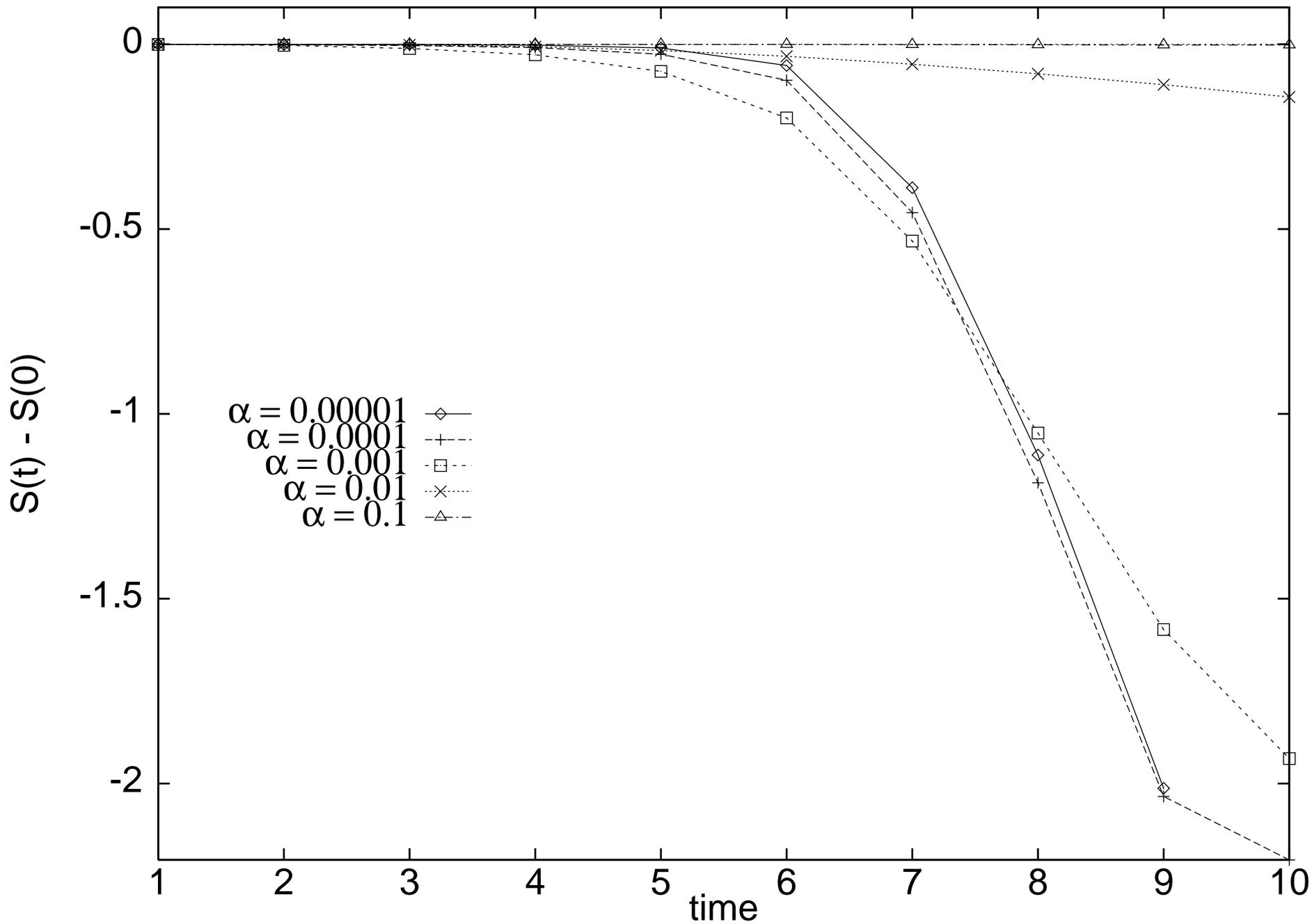
\end{document}